\documentclass[aps,pra,twocolumn,tightenlines,floatfix,showpacs,superscriptaddress,10pt]{revtex4-1}
\usepackage{amsmath}
\usepackage{amsfonts}
\usepackage{amssymb}
\usepackage{fancyvrb}
\usepackage{graphicx}
\usepackage{color}
\usepackage[percent]{overpic}

\renewcommand{\section}[1] {\bigskip \noindent {\large \textbf{#1}}}
\renewcommand{\subsection}[1] {\medskip \mathversion{bold}\noindent\textbf{#1}\mathversion{normal}}

\renewcommand{\section}[1] {}
\renewcommand{\subsection}[1] {}

\begin{document}


\author{Massimiliano Di Ventra}
\affiliation{
	Department of Physics, University of California, San Diego, La Jolla, 92093 CA}

\author{Fabio L. Traversa}
\affiliation{
Department of Physics, University of California, San Diego, La Jolla, 92093 CA}

\author{Igor V. Ovchinnikov}
\affiliation{Electrical Engineering Department, University of California at Los Angeles, Los Angeles, 90095 CA}

\title{Topological field theory and computing with instantons}

\maketitle

\subsection{}\textbf{
Chern-Simons topological field theories (TFTs)~\cite{TQC2,Book} are the only TFTs that have already found application in the description of some exotic strongly-correlated electron systems and the corresponding concept of topological quantum computing~\cite{TQC1,TQC2}. Here, we show that TFTs of another type, specifically the gauge-field-less Witten-type TFTs known as topological sigma models~\cite{Witten2}, describe the recently proposed digital memcomputing machines (DMMs)~\cite{UMM,DMM2} -- engineered dynamical systems with point attractors being the solutions of the corresponding logic circuit that solves a specific task. This result derives from the recent finding that any stochastic differential equation possesses a topological supersymmetry~\cite{Entropy,STS_chapter}, and the realization that the solution search by a DMM proceeds via an instantonic phase. Certain TFT correlators in DMMs then reveal the presence of a transient long-range order both in space and time, associated with the effective breakdown of the topological supersymmetry by instantons. The ensuing non-locality and the low dimensionality of instantons are the physical reasons why DMMs can solve complex problems efficiently, despite their non-quantum character. We exemplify these results with the solution of prime factorization. 
}



Among the physical computing paradigms that have received considerable attention in the past decade, topological quantum computation \cite{TQC1,TQC2} is a very promising form of unconventional computing. It exploits, in an essential way, the topological features of the ground states of some exotic strongly-correlated quantum electron systems, such as p-wave superconductors and fractional quantum Hall systems, to realize computation unencumbered by decoherence and noise \cite{TQC3}. 
As of now, these low-temperature phases of quantum matter are the only known physical realizations of the Chern-Simons topological field theories (TFTs) \cite{TQC2, Book}, the theoretical framework of topological quantum computing. 

There are, however, other types of TFTs known as topological sigma models (TSMs) \cite{Witten2}, a subset of the Witten-type TFTs without gauge fields \cite{Witten1}. These are endowed with a topological supersymmetry leading to the topological character of certain correlators calculated, for example, on \emph{instantons}, namely families of classical solutions of (highly-nonlinear) equations of motion. So far no computational paradigm has been described by TSMs in the same way as topological quantum computing by Chern-Simons TFTs. 

On the other hand, using methods of dynamical systems theory \cite{Ruelle} and stochastic quantization \cite{ParSour}, it was recently shown that all stochastic differential equations (SDEs), that describe all natural and engineered dynamical systems, including all non-quantum computational devices, possess a topological supersymmetry \cite{Entropy,STS_chapter}. The supercharge of this supersymmetry is the exterior derivative on the phase space, and its ubiquitousness in dynamical systems is the algebraic reflection of the preservation of the continuity of the phase space by a continuous-time dynamics. It is the spontaneous breakdown of this topological supersymmetry that encompasses the theoretical essence of such physical phenomena as 1/f noise, turbulence, chaos, etc. \cite{Entropy}. 

It then follows that if we found dynamical systems that employ instantons to compute, we would have a non-quantum computing paradigm that may take advantage of some of the topological features of the corresponding TSM. Such a paradigm would be very compelling from the theoretical point of view, as it would provide yet another solid connection between topology and computing, and desirable from a practical standpoint since dynamical systems can be engineered to operate at any temperature.

In this paper, we show that the recently proposed concept of {\it memcomputing} (computing with and in memory) \cite{diventra13a,UMM,DMM2} is such a computational paradigm. We focus on  {\it digital} memcomputing machines (DMMs), that, being maps of integers into integers, are easily scalable and hence more promising in a practical context \cite{UMM,DMM2}. We show that DMMs solve problems by means of an {\it instantonic phase} in which the topological supersymmetry is effectively broken, giving rise to a dynamical long-range order. 
The long-range order allows logic elements of the machine to correlate to other elements arbitrarily far away, despite its classical character. In addition, the instantonic phase -- contributing to the action with zero measure -- scales at most polynomially with system size, thus making the solution search very efficient. We argue that these are the main physical reasons behind the computational power of DMMs \cite{UMM,DMM2}.

\begin{figure}
	\centering
	\begin{overpic}[width=0.47\textwidth]{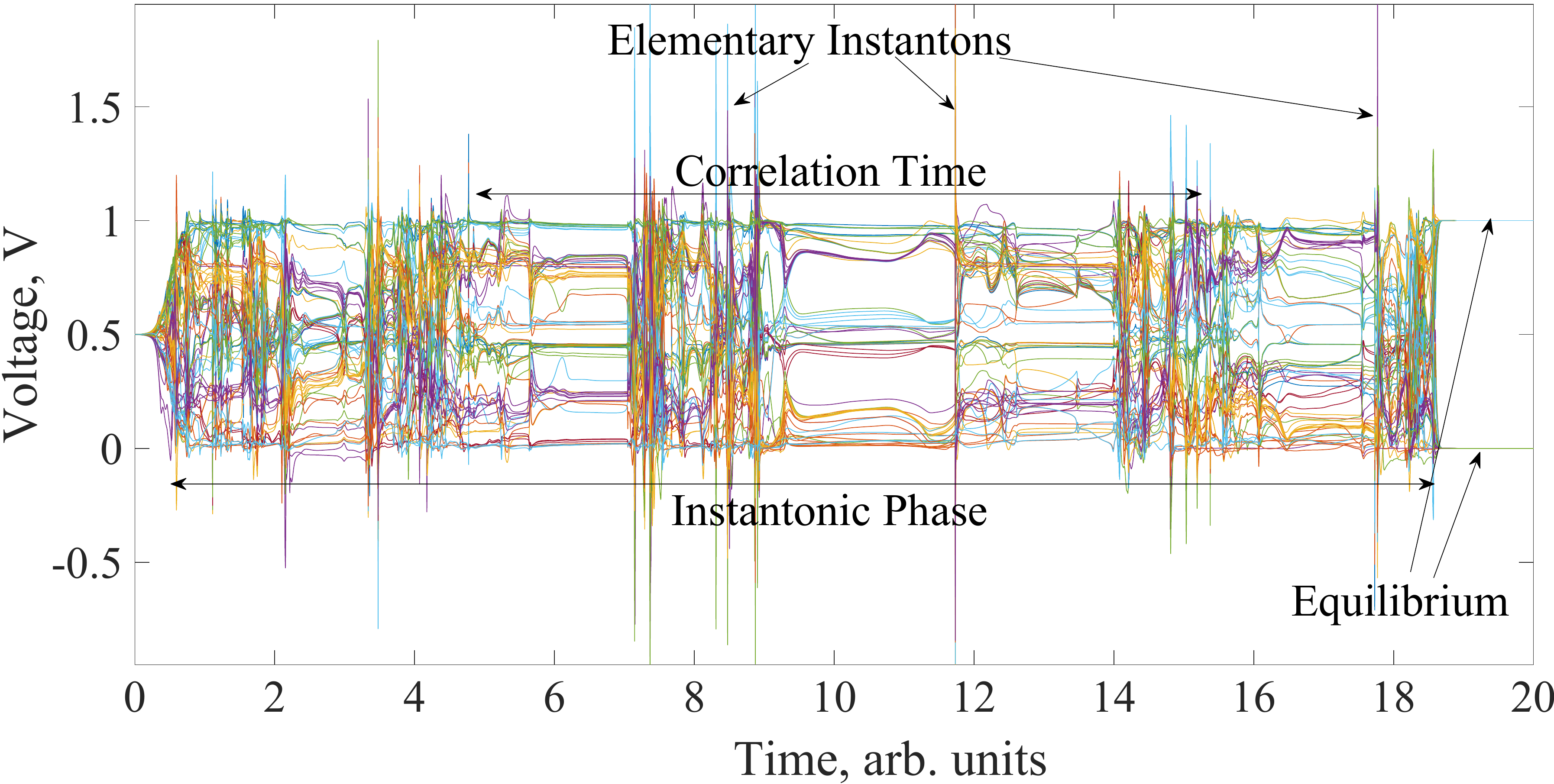}
		\put (0,46) {(a)}
	\end{overpic}
	\par\vspace{.2cm}
	\begin{overpic}[width=0.47\textwidth]{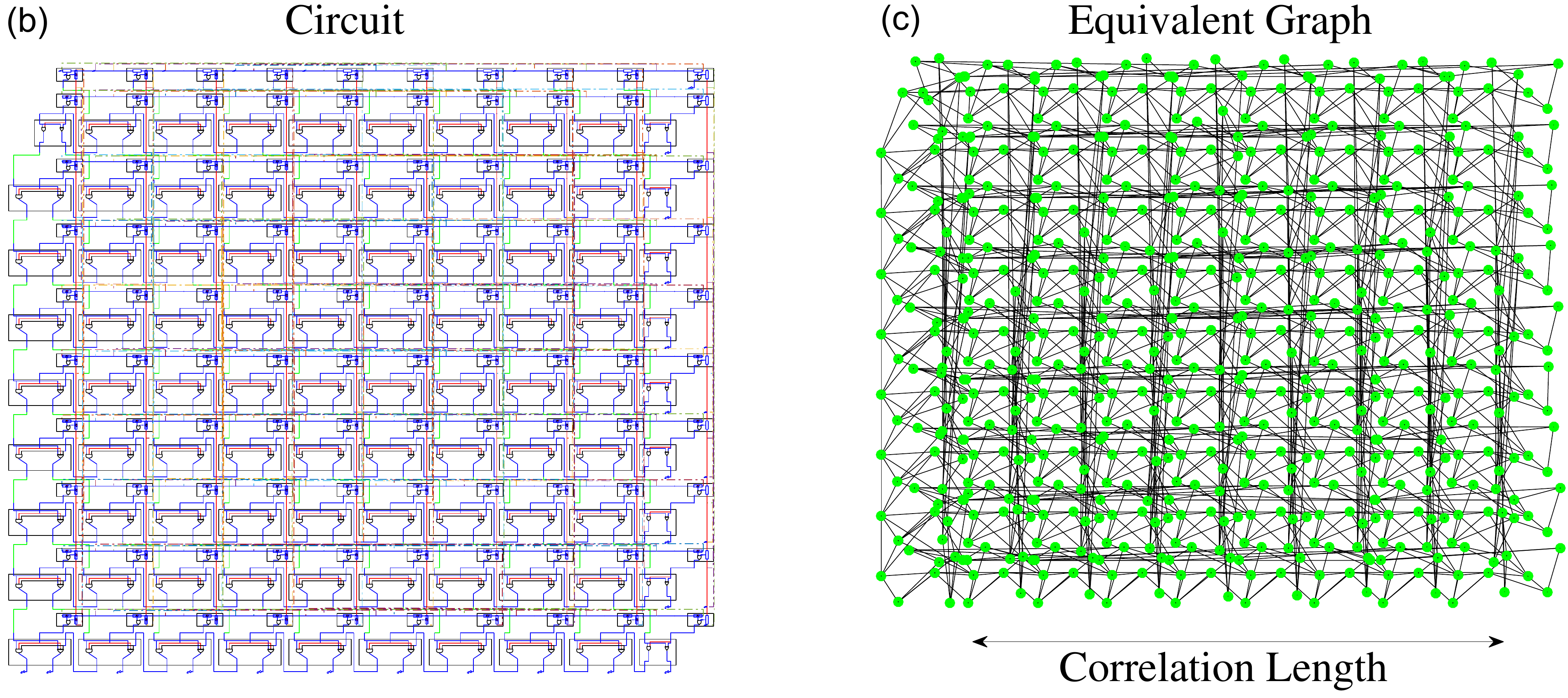}
		\put (0,42) {(b)}
		\put (57,42) {(c)}
    \end{overpic}
	\caption{\label{Figure_1}{\bf Solution of factorization via instantons.} (a) Voltages vs. time at a randomly chosen subset of 70 literals (nodes of the circuit). The instantonic phase consists of elementary instantons. The correlation time defined as the time $\tau$ for which the time correlation $C(\tau)$ of Fig.~\ref{Figure_3} drops by one order of magnitude is highlighted by the double head arrow (see also Methods). (b) Self-organizing logic circuit for the  factorization of a 20-bit number. The number factorized is $497503=499\times 997$. (c) Equivalent undirected graph representing the circuit in (b). The graph representation is used to define a distance between literals (circuit nodes). The correlation length defined as the graph distance for which the time correlation $C(d)$ of Fig.~\ref{Figure_3} drops by one order of magnitude is highlighted by the double head arrow (see also Methods).}  
\end{figure}
	

To exemplify these statements let us start by discussing how a DMM solves, \emph{e.g.}, prime factorization. Consider an integer $n=p\times q$, with $p$ and $q$ some primes. All these integers are represented as bits. Construct a logic circuit that performs the multiplication between $p$ and $q$ to give $n$, \emph{e.g.}, the one in Fig.~\ref{Figure_1}b using AND gates and full adders. This logic circuit is clearly not unique and one may choose any combination of logic gates that satisfies the above problem. We refer the reader to Ref.~\cite{DMM2} for actual details of these logic gates, that, however, are irrelevant for our discussion of their topological properties.
The important point is that the gates we consider, unlike the standard ones, are realized by units that possess time non-locality (memory)~\cite{DMM2}. As a result, they can also work ``in reverse'', in the sense that they are able to {\it adapt} to ''input'' signals from {\it any} terminal, without the distinction between input and output terminals as in conventional logic. In this respect, our circuit is equivalent to an undirected graph (see Fig.~\ref{Figure_1}c) in contrast to the standard boolean circuits that are represented through directed acyclic graphs. 

Now, by switching on the signals (\emph{e.g.}, voltages) that represent the integer $n$ to factorize, the machine attempts to satisfy all the logic propositions at once. This creates a highly-frustrated configuration of metastable logic propositions, with the literals of one gate frustrated by the literals of the neighboring gates. Naively, one then expects that, since a DMM is a non-quantum system, if the input size increases, an exponentially increasing number of resources (both in space and time) is needed to ``sort through'' all the possible states of logical inconsistency. Instead, after a transient that scales at most polynomially with the system size~\cite{DMM2}, and which is a succession of very sudden changes of the literals spanning the {\it entire network} (the spikes of Fig.~\ref{Figure_1}a), DMMs converge to the appropriate equilibrium/solution, thus providing the correct pair of $p$ and $q$ (see Fig.~\ref{Figure_1}a). 

The mathematical proof of these statements has been carried out using functional analysis~\cite{DMM2}. From the physical point of view, this speed-up can be explained by invoking a sort of ``self-organization'', which suggests the presence of a spontaneous long-range order in space and time, through which the logic gates of the DMM communicate and find the shortest path to the solution. We now aim to unravel this order explicitly. 




Let us then start by writing a general DMM as a dynamical system~\cite{DMM2}. All its elements, whether voltages, currents and internal state (memory) variables, that we collectively call $x\in X, X=\mathbb{R}^D$ (with $D$ the dimension of the phase space, $\Omega(X)$) are then some functions of time obeying
\begin{eqnarray}
\dot x(t) = F(x(t))+(2\Theta)^{1/2}e_a(x(t))\xi^a(t)\equiv \mathcal{F}(t), \label{SDE}
\end{eqnarray}
with $F\in TX$ being the flow vector field (on the tangent space $TX$) representing the laws of temporal evolution of $x$, $\xi$ is a Gaussian white noise, $\Theta$ is the intensity of the noise, and $e_a$ are vector fields that represent the coupling of the noise to the system. We can always consider the deterministic limit, $\Theta\to0$, but we include the noise explicitly so that we will determine the stability 
of these machines with no further effort~\footnote{The addition of the noise has the extra mathematical benefit of rendering the stochastic evolution operator elliptic.}. 

Although the actual form of $F$ is very important for the successful operation of a DMM, since it encodes the logical circuit that solves a given problem (\emph{e.g.}, factorization)~\cite{UMM,DMM2}, its explicit form will not be needed for our discussion. What is needed is the knowledge that the only attractors of $F$ are the point attractors representing the solutions of the logic circuit, and $F$ represents a {\it non-linear} dynamical system. As it was demonstrated in Refs.~\cite{UMM,DMM2}, these important properties of $F$ are indeed achievable by fine engineering.

The topological character of Eq.~(\ref{SDE}) is more evident by using its path-integral representation. Following the standard stochastic quantization procedure \cite{Entropy}, we then construct the following functional,
\begin{eqnarray}
{\mathcal W} = \left\langle \iint Dx \delta(\dot x - {\mathcal F})\text{Det}\frac{\delta(\dot x(t) - {\mathcal F}(t))}{\delta x(t')}\right\rangle.\label{PathIntWittenIndex}
\end{eqnarray}
Here, the path integration is over the closed paths in $X$; the infinite-dimensional determinant represents the functional Jacobian of the argument of the preceding $\delta$-functional, and the angled brackets denote the stochastic averaging over all the configurations of the noise. 


To exponentiate the bosonic $\delta$-functional and the fermionic functional determinant in Eq.~(\ref{PathIntWittenIndex}) one can use the standard technique \cite{ParSour,Book_Peskin} of introducing additional fields: the Lagrange multiplier, $B_i$, and the pair of Faddeev-Popov fermionic ghosts, $\chi^i$ and $\bar\chi_i$, with $i$ running over all the variables of the model. After integrating out the noise, Eq.~(\ref{PathIntWittenIndex}) takes the form of the so-called Witten index,
\begin{eqnarray}
{\mathcal W} = \int dx d\chi \hat {\mathcal M}_{tt'}(x\chi,x\chi),\label{Witten_Path}
\end{eqnarray}
where 
\begin{eqnarray}
\hat {\mathcal M}_{tt'}(x\chi,x'\chi') = \iint_{\genfrac{}{}{0pt}{}{\chi(t)=\chi,\chi(t')=\chi'}{x(t)=x,x(t')=x'}} D\Phi \; e^{\{{\mathcal Q},\Psi\}}\label{MathaiQuillen}
\end{eqnarray}
is the most important object in the theory -- the finite-time stochastic evolution operator (SEO) \footnote{The mathematical meaning of the finite-time SEO is the stochastically averaged action, or pullback, induced on the exterior algebra of the phase space by the noise-configuration-dependent SDE-defined diffeomorphisms \cite{Entropy}.}. Path integration in Eq.~(\ref{MathaiQuillen}) is over trajectories connecting the arguments of the SEO, $\Phi=(x,B,\chi,\bar\chi)$ denotes the collection of all the fields, and the operator of the topological supersymmetry (or Becchi-Rouet-Stora-Tyutin (BRST) symmetry) is defined as
\begin{eqnarray*}
\{ {\mathcal Q}, A(\Phi)\} = \int_{t'}^t d\tau \left(\chi^i(\tau)\frac{\delta }{\delta x^i(\tau)}+B_i(\tau)\frac{\delta }{\delta \bar\chi_i(\tau)}\right) A(\Phi), 
\end{eqnarray*}
with $A(\Phi)$ being an arbitrary functional of $\Phi$, and
$\Psi = \int^{t}_{t'} \left(i\hat{\bar\chi}_i(\tau) \dot x^i(\tau) - \bar d(\Phi(\tau))\right)d\tau$ is the so-called gauge fermion, with $\bar d(\Phi) = i\bar\chi_iF^i - \Theta i\bar\chi_ie_a^i \{{\mathcal Q}, i\bar \chi_ke_a^k\}$.

Eq.~(\ref{Witten_Path}) is the standard path-integral representation of a Witten-type (cohomological) TFT -- more specifically a TSM (due to the absence of gauge fields) -- with the action consisting only of a gauge-fixing term \cite{Book}. It is also of a topological character since it depends only on the topology of the phase space, and physically it represents, up to a topological factor, the partition function of the noise \cite{STS_chapter}. 


Now that we have established the TSM representation of DMMs, let us try to understand where their computational power originates from. The objects of interest in this TFT are the instantons, \emph{i.e.}, families of deterministic trajectories,
\begin{eqnarray}
\dot x_{cl}(t,\sigma) = F(x_{cl}(t,\sigma));\;\;\;\;\;\;  x_{cl}(\pm\infty,\sigma)=x_{a,b},\label{instanton}
\end{eqnarray}
that connect two arbitrary critical points of $F$, say $x_b$ and $x_a$. The parameters $\sigma$ are the so-called modulii of instantons, and can be used as their coordinates. For simplicity, we also assume that all the critical points of $F$ are isolated, which is a reasonable assumption in the case of DMMs~\cite{DMM2}. 

As can be seen from Fig.~\ref{Figure_1}a -- which reports the voltages at randomly selected subset of literals of the logic gates in the circuit of Fig.~\ref{Figure_1}b that solves the factorization of a number of 20 bits-- the computation begins at an unstable critical point of $F$, and ends at one of its stable point attractors representing the solutions of the logic circuit (in the example of Fig.~\ref{Figure_1} only one solution is present). The bursts of dynamics in Fig.~\ref{Figure_1}a represent the \emph{elementary instantons} of the composite instantonic phase.  

The coupling between the nonlinear elements of the logic circuit introduces dynamical constraints. In other words, not all logic elements are independent for the ``low-energy'' dynamics. This guarantees that the dimensionality of instantons, say $m$, is considerably smaller than the dimensionality of the phase space, $D$. The operation of a DMM can then be interpreted as the instatonic process that rids of these $m$ unstable, collective variables (logical defects). Each burst signifies that yet another logical defect has left the DMM. The number of these defects, in turn, cannot scale faster than a polynomial of the dimensionality of the ``input'' vector, namely the number of bits representing, say, the number to be factorized. Indeed, one cannot create, say, two unstable variables by flipping only one literal. This is a physical realization of the well-known and general fact that instantons have {\it zero measure} contribution to the action~\cite{Coleman}.

\begin{figure}
\centering
\includegraphics[width=.70\linewidth]{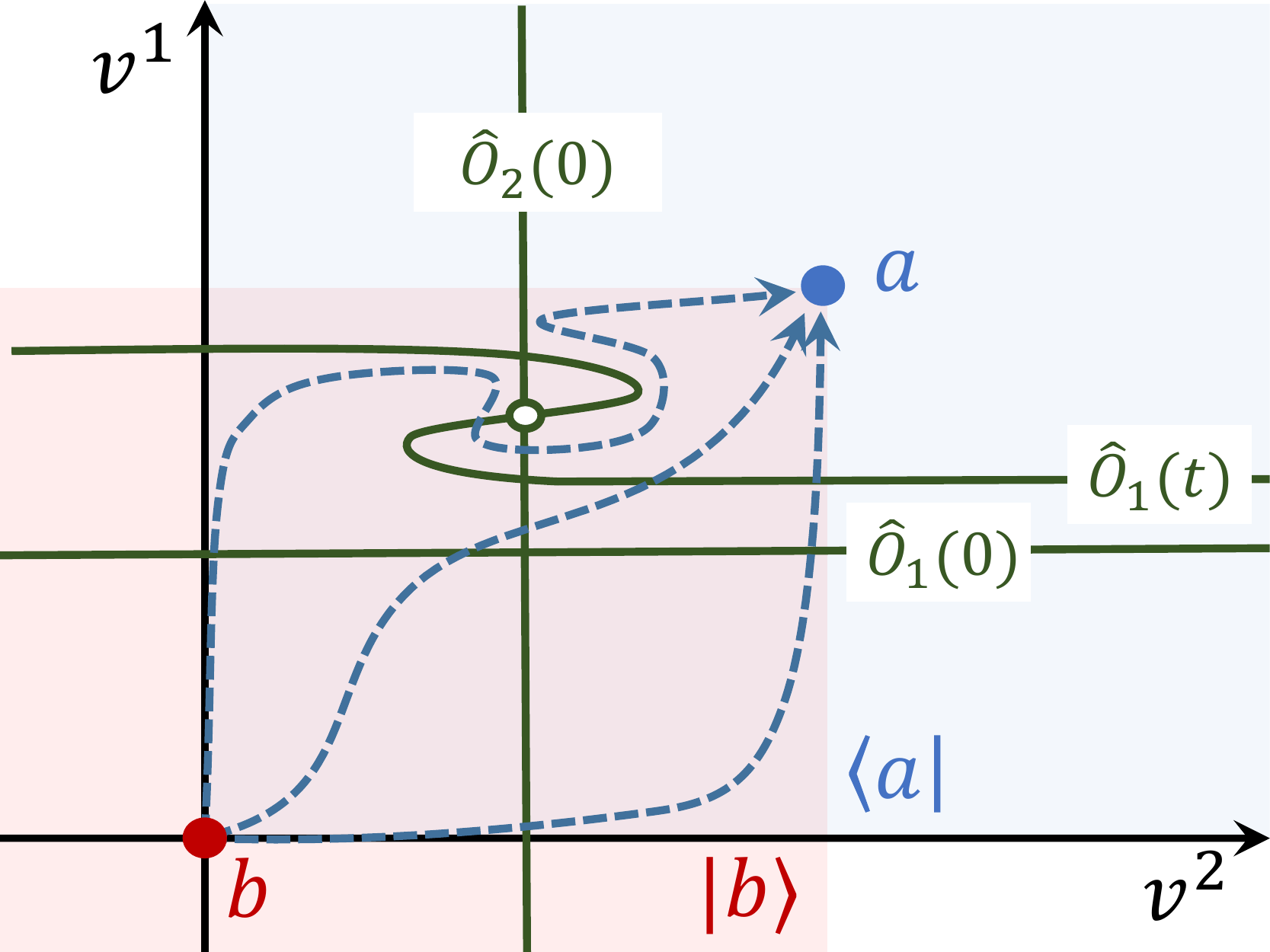}
\caption{\label{Figure_2} {\bf Two-dimensional schematic representation of the instantonic matrix element in Eq.~(\ref{InstantonicME})}. The bra of the $a$-vacuum, $\langle a|$ (blue area), is the Poincar\'e dual of the attraction basin of $a$, and the ket of the $b$-vacuum, $|b\rangle$ (red area), is the repulsion basin of $b$. The instanton, $\langle a|...|b\rangle$, is the overlap that consists of all trajectories (dashed-curved arrows) starting at $b$ and ending at $a$. The "on-time" instantonic matrix element $\langle a|\hat O_1(0)\hat O_2(0)|b\rangle=1$, is the intersection number of the two hyperplanes, $v^{1,2}=1/2$. While the time argument of $\hat O_1(t)$ changes, the intersection number remains the same since new intersections appear in pairs of positive and negative (hollow circle) Jacobians, thus canceling each other in Eq.~(\ref{finalI}).} 
\end{figure}

The TSM representation of DMMs allows us to calculate instantonic matrix elements of the type
\begin{eqnarray}
{\mathcal I} = \iint_{x(\pm\infty)=x_{a,b}}D\Phi \; e^{\{{\mathcal Q},\Psi\}} \prod\nolimits_{i=1}^{m} O_{\alpha_i}(\Phi(t_i)),\label{InstantonicME}
\end{eqnarray}
where the path-integration is over the fluctuations around instantons, $\alpha_i, i=1, \dots,m \ll D$ is a set of literals in the circuit. Let us now choose observables in the form,\footnote{Here we assume that our instanton has the simplest structure possible as given in Fig. \ref{Figure_2}. Namely, it is a ''square'' such that the unstable point $b$ has all the coordinates zero, whereas the stable point $a$ has all variables unity. This assumption simplifies the calculations but it is by no means necessary for our purposes. The supersymmetric theory of stochastics we use here is coordinate-free and for the class of instantons that we consider one can always choose collective coordinates in such a way that it will render the instanton a ''square''.} 
\begin{eqnarray}
O_{\alpha}(\Phi)=\delta(x^{\alpha}-1/2)\chi^{\alpha}.\label{Observ}
\end{eqnarray}
The physical meaning of Eq.~(\ref{Observ}) is to "detect" when the voltages 
on the literals "cross" the value 1/2, either towards the logical 0 or the logical 1, and compensate for the missing ghost (fermion) in each  
unstable direction of the initial state (hence the presence of $\chi^{\alpha}$). For later use, we also note that this observable is $\mathcal Q$-exact, 
\begin{eqnarray}\label{Q-exact}
O_{\alpha}(\Phi(t))=\{\mathcal{Q},\theta(x^{\alpha}(t)-1/2)\},
\end{eqnarray}
with $\theta(x)$ being the Heaviside step function, and the product of any number of $\mathcal Q$-exact operators is $\mathcal Q$-exact itself, $O_{\alpha_1} O_{\alpha_2}...=\{\mathcal Q, \theta(x^{\alpha_1}(t)-1/2) O_{\alpha_2}...\}$, due to the nilpotency of the differentiation by $\mathcal Q$. 

The calculation of $\mathcal I$ is now done in the standard manner. According to the so-called ``localization principle'' (see, \emph{e.g.}, Chap. 9 of Ref.~\cite{Book1}) -- a direct consequence of the topological supersymmetry of the model -- the zeroth order (classical) evaluation of $\mathcal I$ on  instantons is correct to all orders. To perform this calculation one then splits the fields as
\begin{eqnarray}
x(t) = x_{cl}(t,\sigma) + ...,\;\;\;\; 
\chi^i(t) =  \frac{\partial x_{cl}^i(t,\sigma)}{\partial \sigma^j } \chi^{\sigma^{j}} + ...,
\end{eqnarray}
where the dots represent all other fluctuational modes, and $\chi^{\sigma^{j}}$ are the supersymmetric partners of the modulii. By substituting this resolution into Eq.~(\ref{InstantonicME}), and dropping all contributions from orders higher than zero, we get 
\begin{eqnarray}\label{instantmod}
{\mathcal I} &=& \int \prod_{i=1}^{m} d\sigma^i d\chi^{\sigma^{i}} \delta(x^{\alpha_{i}}_{cl}(t_i,\sigma)-1/2)\nonumber
\frac{\partial x_{cl}^{\alpha_i}(t,\sigma)}{\partial \sigma^j } \chi^{\sigma^{j}} 
\\ 
\label{finalI}
&=& \sum_{\sigma_0, x_{cl}^{\alpha_i}(t_i,\sigma_0)=1/2} \text{sign det} \left.\frac{\partial x_{cl}^{\alpha_i}(t_i,\sigma)}{\partial \sigma^j}\right|_{\sigma=\sigma_0}.
\end{eqnarray}
To understand this important result let us first discuss this matrix element in the operator representation. It reads
\begin{eqnarray}\label{Instantonop}
{\mathcal I} = \langle a| \prod\nolimits_{i=1}^m O_{\alpha_i}(\hat \Phi(t_i))|b \rangle,
\end{eqnarray}
where operators are in the Heisenberg representation, $\hat \Phi (t)=\hat {\mathcal M}_{0t}\hat \Phi \hat {\mathcal M}_{t0}$, with $\hat \Phi$ being Schroedinger operators, and the choice of the reference time instant, which is taken to be zero, is irrelevant, and without loss of generality we have assumed that the time moments $t_i$'s are chronologically ordered. The states   
$\langle a |$ and $| b \rangle $ are the bra and ket of the $a-$ and $b-$vacua, namely perturbative supersymmetric ground states associated with the respective critical points. They are the differential forms -- so called Poincar\'e duals \footnote{The Poincar\'e dual of a manifold is a $\delta$-functional distribution on the manifold with fermions in the transverse directions, and a constant function without fermions along the manifold.} -- of the local stable and unstable manifolds for the bra and ket of the vacuum, respectively. This is visualized in Fig. \ref{Figure_2}, where it can be seen that the combination of $\langle a|$ and $|b \rangle$ is the Poincar\'e dual of the instanton (\ref{instanton}).

Observable (\ref{Observ}) can then be recognized as a Poincar\'e dual of the hyperplane $x^{\alpha}=1/2$. Therefore, $\mathcal I$ for all $t_i=0$ can be interpreted as an intersection of a collection of such hyperplanes on the instanton. If one chooses the $\alpha$-set such that during the instanton all the elements "cross" $1/2$, say from $0$ to $1$, then $\mathcal I$ equals unity {\it no matter how far} the logical elements in the $\alpha$-set are separated from each other {\it spatially}. In other words, the instantonic matrix element $\mathcal I$ correlates literals (voltages) arbitrarily far from each other, despite the classical nature of the machines. This spatial long-range order originates from the spatial nonlocality of the collective instantonic variables, which, in turn, takes its origin in the strong nonlinear interactions constituting the DMM logical circuit~\cite{DMM2}. 

Furthermore, as schematically shown in Fig.~\ref{Figure_2}, it is also clear why $\mathcal I$ is independent of time variables: if the time argument of the observables in Eq.~(\ref{Instantonop}) changes, pairs of positive and negative Jacobians appear, canceling each other in Eq.~(\ref{finalI}). This demonstrates that the transient dynamics of DMMs has also a {\it temporal long-range order}. Indeed, at the initial point of the evolution, the system has unstable variables (instanton modulii) and thus the trajectory is highly sensitive to the external influence and/or initial conditions. This temporal long-range order can be interpreted as the sort of dynamical long-range order in ordinary chaotic dynamical systems, that have topological supersymmetry broken spontaneously~\cite{STS_chapter}. 

More precisely, when the topological supersymmetry is spontaneously broken one can calculate non-zero expectation values of $\mathcal Q$-exact operators on the \emph{global} (non-supersymmetric) ground state(s), $\langle g|\{{\mathcal Q}, A \}|g\rangle\ne 0$. Similarly, the instantonic matrix element is that of a $\mathcal Q$-exact operator because the observable we have chosen is $\mathcal Q$-exact (c.f. Eq.~(\ref{Q-exact})). However, the analogy between the temporal long-range order in DMMs and the dynamical long-range order in ordinary chaotic dynamical systems is only qualitative for two reasons. First, ${\mathcal I}$ is not a diagonal matrix element. Second, the two different vacua involved in $\mathcal I$ are local, not global ground states of the model.

\begin{figure}
	\centering
		\begin{overpic}[width=0.47\textwidth]{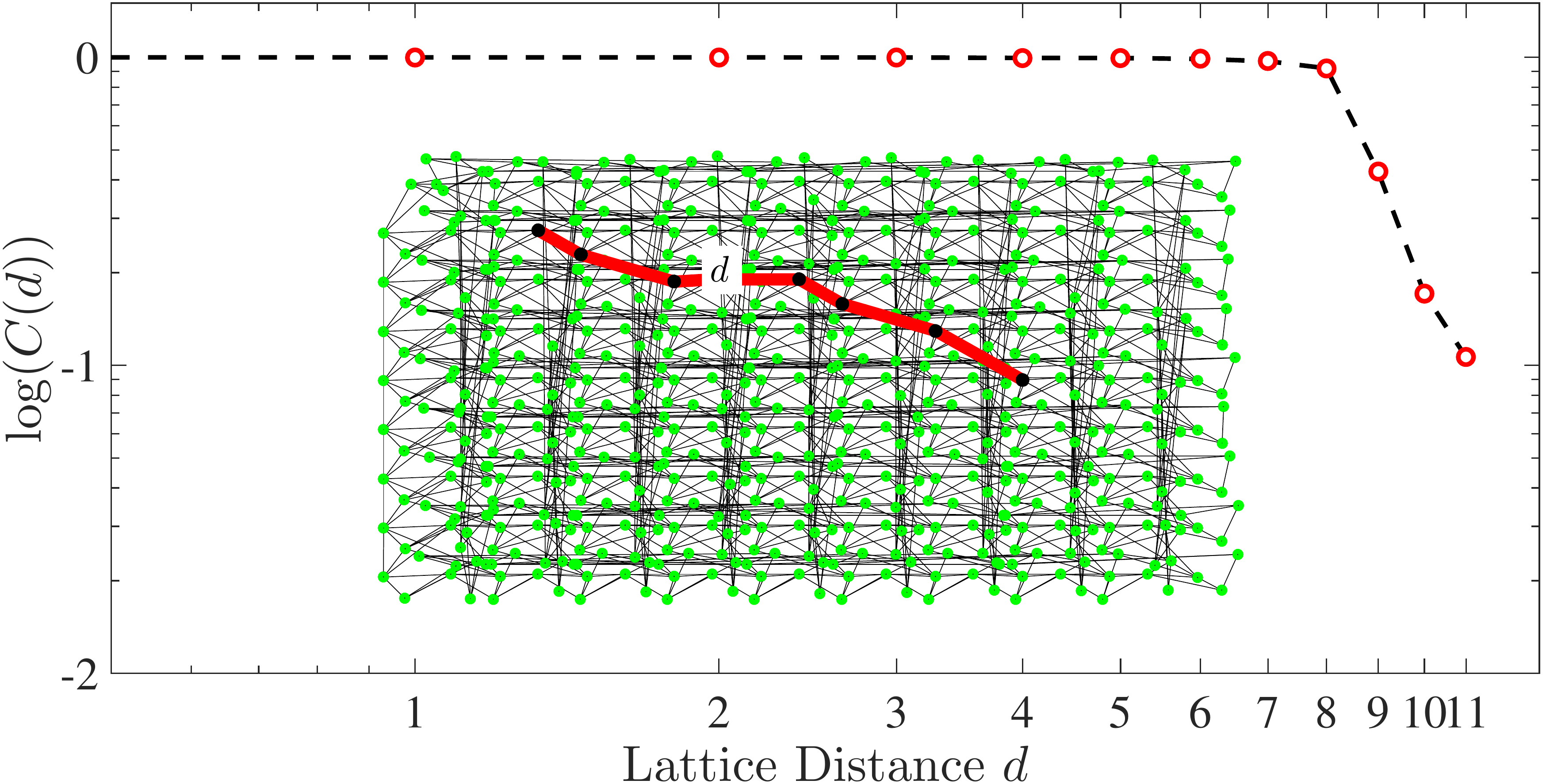}
			\put (0,47) {(a)}
		\end{overpic}
\par\vspace{.2cm}
	\begin{overpic}[width=0.47\textwidth]{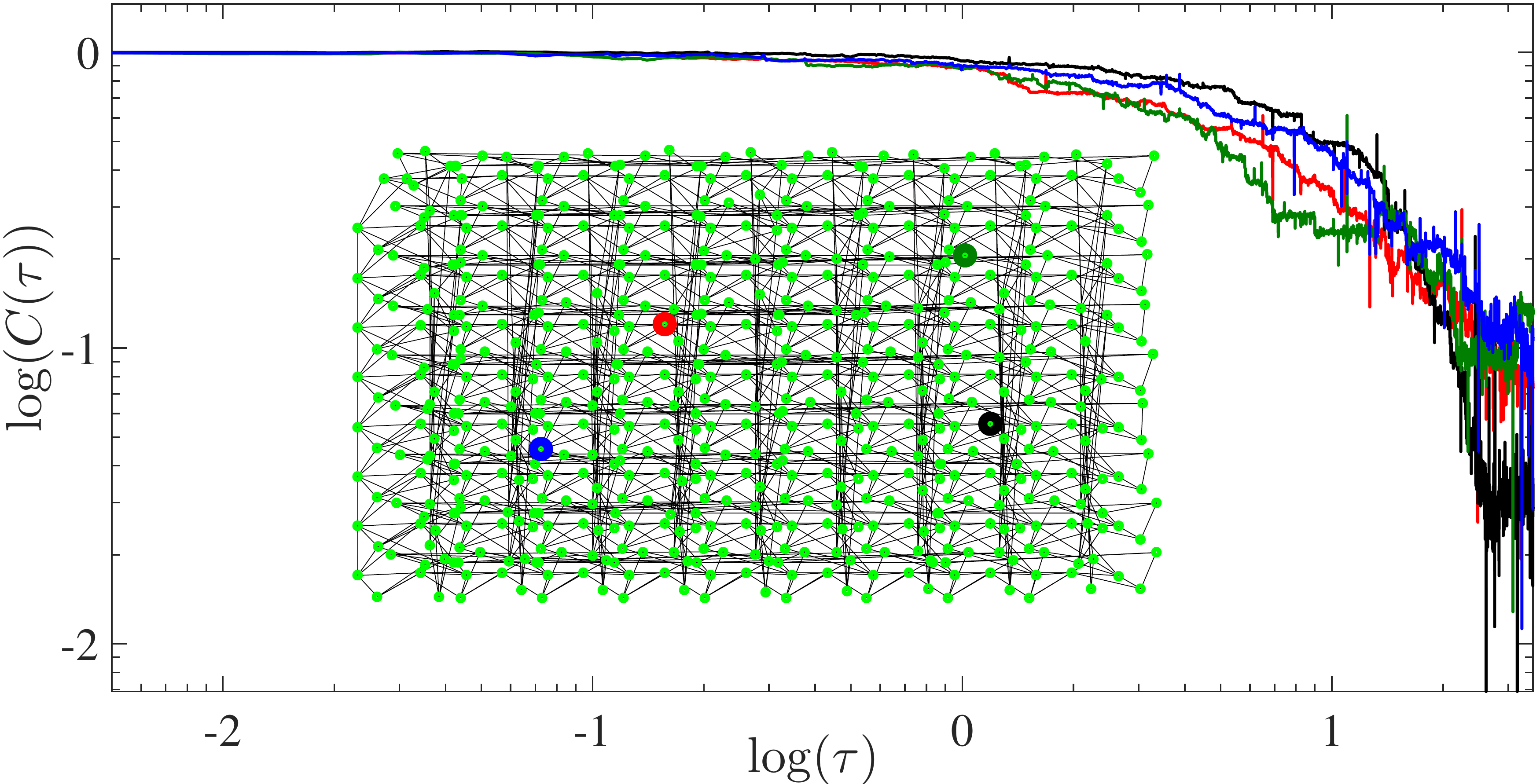}
		\put (0,49) {(b)}
	\end{overpic}	
	\caption{\label{Figure_3} {\bf Long-range spatial and temporal order.} (a) Normalized spatial correlation $C(d)$ where $d$ is graph distance sketched in the inset of the figure. (b) Normalized temporal correlations $C(\tau)$ of four random selected literals indicated in the inset of the figure (see Methods).} 
\end{figure}



Note also that the long-range order is limited only to the transient time when the DMM needs to connect the initial state and the final solution state(s). We show this explicitly for the spatial and temporal correlations in Fig.~\ref{Figure_3} for the DMM of Fig.~\ref{Figure_1} (see Eqs.~(\ref{Cd}) and~(\ref{Ctau}) in the Methods). The numerical results of Fig.~\ref{Figure_3} confirm our theoretical analysis: the correlations are essentially constant and drop off only when the distance is comparable with the size of the circuit (spatial correlations -- the maximum distance of the equivalent graph in this case is 11 and the correlation length defined in Fig.~\ref{Figure_1} is 9), or when the time difference exceeds the bound of the instantonic phase (temporal correlations).
 



In conclusion, we have shown that the recently suggested digital memcomputing machines -- that employ memory to both compute and store information, and map integers into integers --~\cite{DMM2} can be characterized as physical systems described by topological sigma models. This class of theories allows the calculation of certain topological invariants on instantons. We have shown that the dynamics of memcomputing machines proceed through an instantonic phase that connects the initial supersymmetric state and the final (supersymmetric) solution attractor(s). During this instantonic phase long-range order develops both in space and time, guaranteeing that, despite their non-quantum character, these machines can correlate logic gates {\it anywhere} in the circuit. 
This long-range spatial correlation is a direct consequence of the strong non-linear interactions in the system which reduce the dimensionality of the 
instantonic sector considerably, compared to the dimension of the phase space. In addition, the infinite memory of the time order implies that the machine can choose several paths (if they exist) to solve the given problem with equal efficiency. 

We have exemplified these results with prime factorization, but similar considerations apply to all other complex problems that can be formulated in Boolean form~\cite{DMM2}. Finally, the topological character of these machines implies that they are robust against noise and structural disorder, up to strengths that do not affect the topology of the phase space. Since digital memcomputing machines are classical objects that operate at room temperature, they are easy to build with available materials and devices. Our work then opens up the possibility to realize some sort of topological computing on non-quantum engineered dynamical systems that employ instantons to compute. It also provides an example of application of topological sigma models to a realistic physical system. 

\section{Methods}

{\bigskip \noindent {\large \textbf{Methods}}} 

\medskip \noindent The correlations reported in Fig.~\ref{Figure_3} have been evaluated on an ensemble of 2000 different initial conditions of Eq.~\eqref{SDE} chosen at random with uniform distribution as follows. The 
normalized spatial correlations (at a time $t$ in the middle of the instantonic phase) are 
\begin{equation}
C(d)=\max_{(\alpha,\beta)\in \Sigma_d} \frac{E^2\{v_\alpha(t),v_\beta(t)\}}{\sqrt{E^2\{v_\alpha(t)\}E^2\{v_\beta(t)\}}}\,, 
\label{Cd}
\end{equation}\\
where $E^2\{a,b\}=E\{ab\}-E\{a\}E\{b\}$ and $E^2\{a\}=E^2\{a,a\}=E\{a^2\}-E\{a\}^2$, being $E\{a\}$ the average of a function $a$ made on the ensemble; $v_{\alpha,\beta} $ are the voltages at the literals $\alpha$ and $\beta$, and $\Sigma_d$ is the set of all pairs $(\alpha,\beta)$ such that $dis(v_\alpha,v_\beta)=d$ where $dis$ is the graph distance defined as the shortest path between two nodes (literals) in the undirected, unweighted graph of Fig.~\ref{Figure_1}c. 
The normalized temporal correlations at fixed literal position are computed as
\begin{equation}
C(\tau)=\frac{E^2\{v_\alpha(t),v_\alpha(t+\tau)\}}{E^2\{v_\alpha(t)\}} 
\label{Ctau}
\end{equation}
for a time $t$ in the middle of the instantonic phase, and $\tau$ an arbitrary interval of time.


{\bigskip \noindent {\large \textbf{Acknowledgments}}}

\medskip \noindent  M.D. and F.L.T. acknowledge partial support from the Center for Memory Recording Research at UCSD.

\medskip \noindent 
Correspondence and requests for material should be addressed to M.D. (diventra@physics.ucsd.edu), F.L.T. (ftraversa@physics.ucsd.edu), or I.V.O. (iovchinnikov@ucla.edu).




\bibliographystyle{naturemag}
\bibliography{SUSYref}

\end{document}